\documentclass[preprint,5p]{elsarticle}
\usepackage{amssymb}
\usepackage{amsmath}
\usepackage{hyperref}
\usepackage{epstopdf}
\hypersetup{bookmarks,colorlinks}

\providecommand{\wn}{\text{cm}^{-1}}

\begin{document}

\title{Near-infrared LIF spectroscopy of HfF}

\author[jila,boulder]{M. Grau}
\ead{graum@jila.colorado.edu}
\author[michigan]{A.E. Leanhardt}
\author[jila,boulder]{H. Loh}
\author[jila,boulder]{L.C. Sinclair}
\author[jila,boulder]{R.P. Stutz}
\author[jila,boulder]{T.S. Yahn}
\author[jila,boulder]{E.A. Cornell\corref{cor1}}
\ead{ecornell@jilau1.Colorado.EDU}

\cortext[cor1]{Corresponding author}

\address[jila]{JILA, National Institute of Standards and Technology and University of Colorado, Boulder, CO 80309-0440, USA}
\address[boulder]{Department of Physics, University of Colorado, Boulder, CO 80309-0440, USA}
\address[michigan]{Department of Physics, University of Michigan, Ann Arbor, MI 48109-1040, USA}

\begin{abstract}
The molecular ion HfF$^+$ is the chosen species for a JILA experiment to measure the electron electric dipole moment (eEDM). Detailed knowledge of the spectrum of HfF is crucial  to prepare HfF$^+$ in a state suitable for performing an eEDM measurement\cite{Leanhardt}. We investigated the near-infrared electronic spectrum of HfF using laser-induced fluorescence (LIF) of a supersonic  molecular beam. We discovered eight unreported bands, and assign each of them unambiguously, four to vibrational bands belonging to the transition $[13.8]0.5 \leftarrow X1.5$, and four to vibrational bands belonging to the transition $[14.2]1.5 \leftarrow X1.5$. Additionally, we report an improved measurement of vibrational spacing of the ground state, as well as anharmonicity $\omega_e x_e$.
\end{abstract}


\maketitle

\section{Introduction}
HfF$^{+}$ is one of the candidate species for experiments to measure the electron electric dipole moment (eEDM). In particular, the long-lived metastable $^{3}\Delta_1$ state of the ion possesses both a long coherence time\cite{Meyer,Barker} and a strong effective electric field\cite{Meyer,Titov}, offering enhanced sensitivity to an eEDM signal of perhaps as many as two orders of magnitude\cite{Leanhardt} over the current experimental limit\cite{hinds}. The degree to which an electron possesses an electric dipole moment is deeply connected to the phenomenon of charge-parity violation in particle physics. A more sensitive measurement of the eEDM will stand to constrain the parameters of various so-called Beyond Standard Model theories of high energy physics.

A crucial aspect of the eEDM measurement is preparing a sample of HfF$^{+}$ in the state with the highest eEDM sensitivity, in a given  magnetic sublevel of the $J=1$ manifold of $^3\Delta_1$ $v=0$. The proposed mechanism for this is two-color autoionization directly into the science state\cite{autoionization}. To this end, we require detailed knowledge of the spectrum of HfF to identify intermediate states for the autoionization process.

Spectroscopy of HfF has previously been done by Adam et al.\cite{Adam}. In addition to the present work, we have done spectroscopy on HfF using resonantly-enhanced multi-photon ionization\cite{rempi}. Also of relevance is the work done by Barker et al.\cite{Barker}, which is the first spectroscopy of the low energy states of HfF$^{+}$, as well as other work by our group investigating the higher energy states of HfF$^+$\cite{Cossel,Sinclair}.

\section{Experiment}
We use a supersonic beam apparatus similar  to that which is described in Ref. \citenum{autoionization}. We hold a mixture of 99\%Ar +  1\%SF$_6$ at 100 psi backing pressure. A pulsed valve opens for 140 $\mu$s and the gas undergoes expansion, achieving supersonic velocities after entering the ablation chamber. A 30 mJ pulse of focused light at 1064 nm from an Nd:YAG laser ablates material from a Hf target rod $0.14$ in. in diameter. The ablation plume is entrained in the supersonic beam and chemically reacts to produce HfF, HfF$^{+}$, and other products. The molecular beam passes through a beam skimmer 11 cm from the point of ablation, and then passes through a second skimmer 9.6 cm further downstream. The diameters of the first and second skimmers are 3 mm and 2 mm respectively, and the region between the skimmers is differentially pumped. At a distance of 8.5 cm downstream from the second skimmer the molecules are interrogated using laser-induced fluorescence (LIF). A 130 mJ pulse of light, with a linewidth of 0.04 $\wn$, from a tunable dye laser excites the molecules, which then emit fluorescence which is collected with a gold-coated spherical mirror and focused with an off-axis parabolic mirror onto an R3896 Hamamatsu photomultiplier tube (PMT).  The fluorescence lifetime is long compared to the transit time of 15 $\mu$s. In order to temporally gate the prompt scattered light of the fluorescence laser, there is a 6 $\mu$s delay between the laser firing and the beginning of the 9 $\mu$s window where the PMT is active. The wavelength range of our scans was covered by a single dye, LDS722.

\begin{figure*}
\centerline{\includegraphics[height=2.5in]{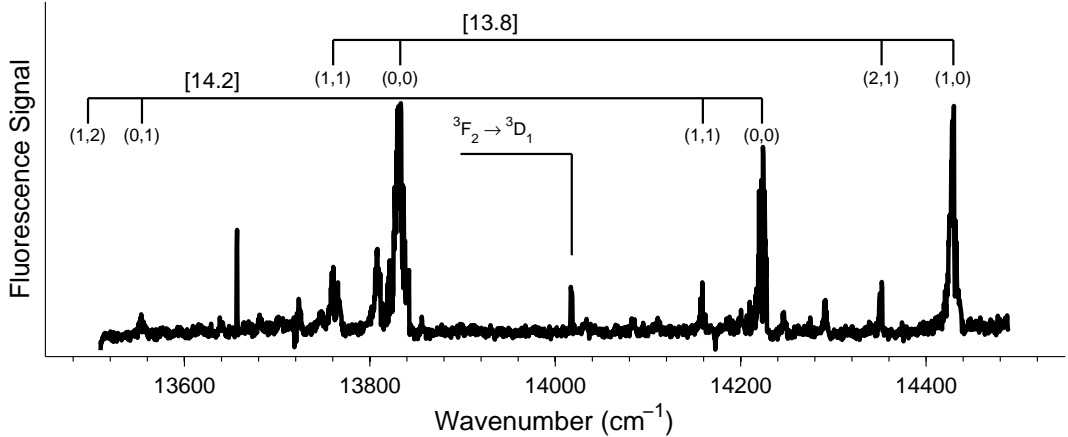}}
\caption{Low resolution survey spectroscopy, with the positions of the eight vibrational bands reported in Table \ref{fit_table} and the Hf $^{3}\text{F}_{2} \to ^{3}\!\!\text{D}_{1}$ transition labeled.}
\label{survey}
\end{figure*}

For the survey spectroscopy in Fig. \ref{survey}, the dye laser is calibrated to Argon lines in an optogalvanic cell, and the laser frequency is determined by interpolating the laser grating position between measurements made with a wavemeter at the beginning and end of the scan. For the high-resolution spectroscopy of the individual bands, for example see Fig. \ref{14223}, the laser frequency is measured at each point in the scan with two separate wavemeters, one with a high absolute accuracy and the other with high resolution, specified to be 10 MHz. The high-resolution wavemeter is calibrated daily to an external-cavity diode laser locked to a $^{87}$Rb transition at 384.227982 THz.

\section{Results and discussion}

\begin{figure}[h!]
\includegraphics{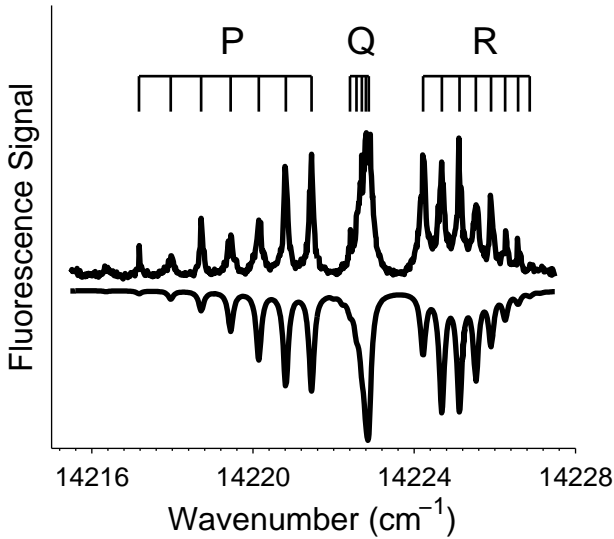}
\caption{The $[14.2]1.5-X1.5 (0,0)$ band of HfF with ground and excited state fit to the Hamiltonian in Eq. \ref{omega}. This demonstrates a $\Delta v=0$ transition with an $\Omega'=3/2$ upper state. The fit is inverted and displayed under the data.}
\label{14223}
\end{figure}

From the survey scan we chose seven vibrational bands to acquire at higher resolution, and following preliminary analysis we looked for and subsequently found an eighth band outside of the original survey range.

In the eight high-resolution scans we see four bands with unresolved isotope splitting, consistent with $\Delta v=0$, and visibly distinct P, Q, and R branches. Of these bands, on the basis of the presence or absence of certain low J lines, two are clearly $(\Omega'=3/2)\leftarrow(\Omega''=3/2)$, and two are clearly $(\Omega'=1/2)\leftarrow(\Omega''=3/2)$ with considerable $\Omega$-doubling. Based on values of the rotational constants, band origin $\nu$, and isotope shifts, we are able to assign these four bands and the remaining four less well-resolved bands to a common lower electronic state, with rotational constants in good agreement with the X1.5 state identified in Ref. \citenum{Adam}. The upper state of these bands we assign  to one of two newly identified electronic states: a state with $\Omega'=3/2$ at 14223 $\wn$, which we label $[14.2]1.5$, and a state with $\Omega'=1/2$ at 13833 $\wn$, which we label $[13.8]0.5$.

On each of the bands acquired at high resolution we perform a non-linear least squares fit to a vibrational band contour. The doubling for all $\Omega=3/2$ levels is unresolved. For the lower states of all eight bands and for the upper states of the bands at 13496, 13554, 14159, and 14223 $\wn$ we fit using the following Hamiltonian:
\begin{align}
F(J) = T + B J (J+1).
\label{omega}
\end{align}
The bands at 13760, 13833, 14352, and 14429 $\wn$ we fit to an upper state using the Hamiltonian\cite{BandC},
\begin{align}
F_a(N) &= T + B N(N+1) - \frac{\gamma}{2}(N+1), \notag\\
F_b(N) &= T + B N(N+1) + \frac{\gamma}{2}N,
\label{sigma}
\end{align}
with selection rules for the P, Q, and R branches:
\begin{align}
P_a &: \left(N'-\frac{1}{2}\right) - J'' = -1, \notag\\
P_b &: \left(N'+\frac{1}{2}\right) - J'' = -1, \notag\\
Q_a &: \left(N'-\frac{1}{2}\right) - J'' = 0, \notag\\
Q_b &: \left(N'+\frac{1}{2}\right) - J'' = 0, \notag\\
R_a &: \left(N'-\frac{1}{2}\right) - J'' = +1, \notag\\
R_b &: \left(N'+\frac{1}{2}\right) - J'' = +1.
\end{align}
The absolute parity of the excited state could not be determined, so following the example in Ref. \citenum{Adam}, we label related pairs as $a/b$ instead of $e/f$ as determined by parity. We do not observe a sufficient number of rotational lines within each band to be able to fit a non-zero value to the higher order rotational constant $D$, as such it is not included in our fitting models. 

\begin{figure}
\includegraphics{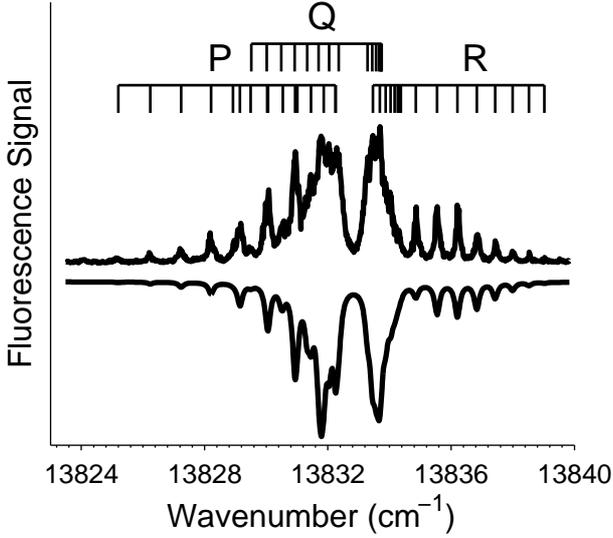}
\caption{The $[13.8]0.5-X1.5 (0,0)$ band of HfF with excited state fit to the Hamiltonian in Eq. \ref{sigma}. This spectrum is an example of a $\Delta v=0$ transition with an $\Omega'=1/2$ upper state with $\Omega$-doubling. The fit is inverted and displayed under the data.}
\label{13832}
\end{figure}

\begin{figure}[h!]
\includegraphics[width=0.5\textwidth]{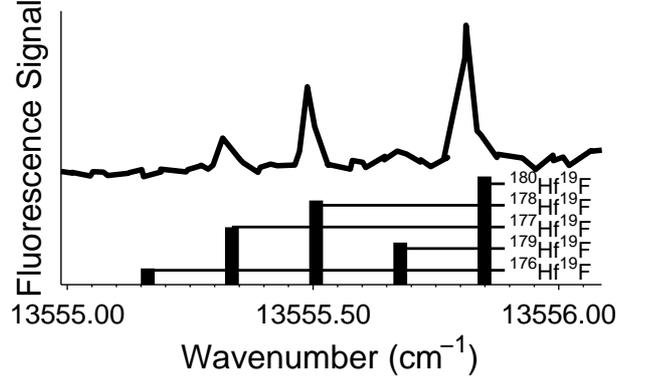}
\caption{A relatively uncluttered portion of the $[14.2]1.5-X1.5 (1,2)$ band highlighting the R(1.5) line, where the 3 most abundant isotopologs of HfF are distinctly visible. Plotted for comparison are the predicted positions of these lines calculated from the reduced mass of each species and the relative intensities expected from natural abundance of each Hf isotope. For low $J$ lines, the intensity of fluorescence peaks from the $^{177}$Hf$^{19}$F and $^{179}$Hf$^{19}$F isotopologs may be diluted due to hafnium hyperfine structure.}
\label{isotopes}
\end{figure}

\begin{figure}[h!]
\includegraphics{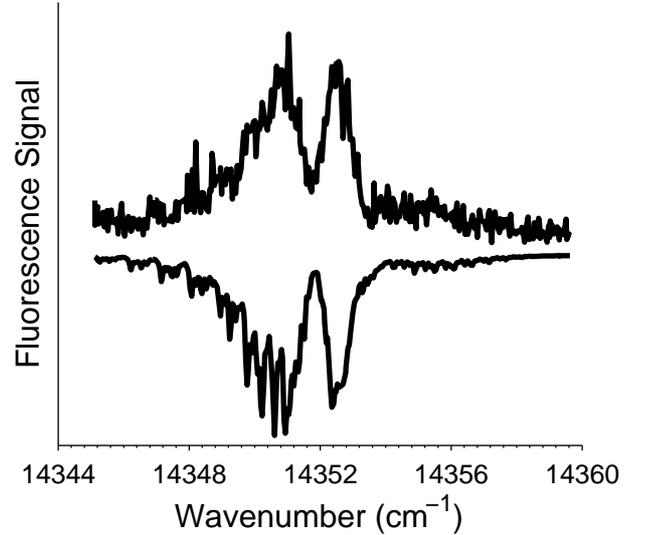}
\caption{The $[13.8]0.5-X1.5 (2,1)$ band of HfF with excited state fit to the Hamiltonian in Eq. \ref{sigma}. The fit is inverted and displayed under the data. For bands of lower quality such as this, our fit only includes $^{\bar{n}}\nu$, $\eta$, overall intensity and rotational temperature as adjustable parameters. Rotational constants are determined from fits to other bands.}
\label{14429}
\end{figure}

\begin{table*}
\renewcommand\tabcolsep{10pt} 
\begin{tabular}{l l l l l l l l l}
\hline \hline										
$^{\bar{n}}\nu (\wn)$&$\Omega'$&$\Omega''$&$^{\bar{n}}B' (\wn)$&$^{\bar{n}}B'' (\wn)$&$\eta (\wn/\text{amu})$&$v'$&$v''$&$\gamma (\wn)$			\\
\hline											
13496.018(8)	&1.5&1.5&0.2673		&0.2791(12)		&-0.176(4)		&1&2		&-			\\
13554.287(4)	&1.5&1.5&0.2693		&0.2822		&-0.172(2)		&0&1		&-			\\
14159.100(4)	&1.5&1.5&0.2673(8)	&0.2822(10)		&0			&1&1		&-			\\
14222.928(6)	&1.5&1.5&0.2693(4)	&0.2839(4)		&0			&0&0		&-			\\
13760.51(8)		&0.5&1.5&0.2681(8) 	&0.2822		&0			&1&1		&0.070		\\
13832.846(8)	&0.5&1.5&0.2676(8)	&0.2833(8)		&0			&0&0		&0.070(4)		\\
14351.78(6)		&0.5&1.5&0.2690(20)	&0.2822		&0.18(4)		&2&1		&0.070		\\
14429.3(1)		&0.5&1.5&0.2681		&0.2838		&0.20(6)		&1&0		&0.070		\\
\hline											
\hline								
\end{tabular}	
\caption{Molecular constants of vibrational bands of HfF, derived from global non-linear least squares fits. The errors determined by bootstrapping the residuals of the non-linear fits  are reported in parentheses as one standard deviation of the last digit. Values reported without uncertainties are held constant in their respective fitting routine. As described in the text, $\eta$ is the per-amu isotope splitting parameter,  and $^{\bar{n}}\nu$ is the band origin of fictitious isotopolog $^{\bar{n}}$Hf$^{19}$F. The fit value for $^{n}B$ is $^{\bar{n}}\!B (^{\bar{n}}\mu/^{n}\mu)$, and the fit value for $^{n}\nu$ is $^{\bar{n}}\!\nu + \eta(\bar{n}-n)$}
\label{fit_table}	
\end{table*}

There are five isotopologs of HfF, labeled $^{n}$Hf$^{19}$F, (with $n = 180, 179, 178, 177, $and $ 176$) each with corresponding reduced mass $^{n}\mu$, and respective relative abundances 35.1\%, 13.6\%, 27.3\%, 18.6\%, and 5.26\%\cite{iupac}. For $\Delta v=0$ transitions (for instance, as in Figs. \ref{14223} and \ref{13832}) the isotope structure is not resolved in our spectra as it is less than the 0.04 $\wn$ linewidth of the LIF laser. For $\Delta v = +1$ and $\Delta v = -1$ transitions the isotope spacing is in principle resolvable, but with the exception of notably sparse regions of the spectrum (Fig. \ref{isotopes}) we are unable to assign individual rotational lines due to the overlapping spectrum of each isotopolog (as in Fig. \ref{14429}).

In light of this difficulty we adopt the following fitting strategy. We assume that the rotational constants of each isotopolog scale reliably with reduced mass $^{n}\mu$:
\begin{align}
^{n}\!B = ^{\bar{n}}\!B (^{\bar{n}}\mu/^{n}\mu),
\end{align}
where $\bar{n} = 178.494$ labels the abundance-weighted mass average of Hf, and the quantities $^{\bar{n}}\mu = 17.171$ amu and $^{\bar{n}}B$ are the reduced mass and rotational constant of this fictitious mass-averaged isotopolog. Similarly, we take the band origins of each isotopolog to be distributed about their averaged value as follows:
\begin{align}
^{n}\nu = ^{\bar{n}}\!\nu + \eta(\bar{n}-n),
\end{align}
where $\eta$ is the per-amu isotope splitting parameter. The reduced mass of any isotopolog differs from the average mass by a fractional amount of no more than 0.2\%,  justifying this linear expansion at the level of accuracy we aspire to. To the extent the isotope shift is dominated by the vibrational contribution, we expect $\eta$ to be
\begin{align}
\eta = \frac{d}{d \epsilon} \left. \sqrt{\frac{^{\bar{n}+\epsilon}\mu}{^{\bar{n}}\mu}} \right|_{\epsilon=0}\Delta V,
\end{align}
where $\Delta V$ is the difference in vibrational energy between the upper and lower states. It is assumed that at our level of accuracy we are not able to observe the isotope shift in the spin-doubling parameter $\gamma$.

\begin{table*}
\renewcommand\tabcolsep{20pt} 
\begin{tabular}{l |l| l l l l}
\hline \hline										
State			&$v$	&$^{\bar{n}}T (\wn)$		&$^{\bar{n}}B (\wn)$		&$\gamma (\wn)$	&$\Delta G_{v+1/2} (\wn)$	\\	
\hline	
X1.5			&0	&0			&0.2838(4)	&-			&668.641(7)		\\
			&1	&668.641(7)		&0.2822(10)	&-			&663.082(9)		\\
			&2	&663.082(9)		&0.2791(12)	&-			&			\\
$[13.8]0.5$		&0	&13832.846(8)	&0.2676(8)	&0.070(4)		&596.5(1)		\\
			&1	&14429.3(1)		&0.2681(8)	&0.070(4)*		&591.3(1)		\\
			&2	&15020.42(6)	&0.2690(20)	&0.070(4)*		&			\\							
$[14.2]1.5$		&0	&14222.928(6)	&0.2693(4)	&-			&604.813(6) 	\\
			&1	&14827.741(8)	&0.2673(8)	&-			&			\\	
\hline											
\hline								
\end{tabular}	
\caption{Molecular constants for the observed vibrational states of HfF. Values indicated by a dash are where no $\gamma$ parameter is included in the Hamiltonian describing that state. Values noted by an asterisk are for states where the data is not well enough resolved to obtain a precise value of $\gamma$ from a fit. At this level of accuracy, $\gamma$ is not likely to depend on the vibrational level, and thus we present the measured value of $\gamma$ for the $[13.8]0.5$ $v=0$ state as a suggested value for the $v=1$ and $v=2$ levels of the same state.}
\label{mol_table}	
\end{table*}

We use different fitting procedures for $\Delta v=0$ bands compared to $\Delta v = \pm1$ bands. For $\Delta v=0$ bands we assume $\eta = 0$ and fit values of $^{\bar{n}}\nu, ^{\bar{n}}\!B', ^{\bar{n}}\!B''$, temperature, instrumental resolution, and for the $\Omega'=1/2$ bands also $\gamma$. When the same vibrational levels are involved in bands with $\Delta v\not=0$, we use previously determined values of $^{\bar{n}}\!B', ^{\bar{n}}\!B''$, and $\gamma$, and fit a contour in order to extract $^{\bar{n}}\nu$ and $\eta$ only. The exception to this procedure was for the $v=2$ levels of the X1.5 and $[13.8]0.5$ states. For these states we had no alternative but to extract the corresponding values of $B$ from the less well resolved bands. We calculate tentative, purely statistical, one-sigma error estimates using a bootstrap method on the resampled residuals\cite{bootstrap}.  This method assumes the fitting model is perfect, and ignores the possibility of, e.g., nonthermal rotational populations or other deviations from H\"{o}nl-London intensity predictions.  We treat these estimates as indicative of the relative sizes of the various error bars. From inconsistencies observed in values of parameters determined redundantly by fits to different bands, we find that a global doubling of each error bar is appropriate. A summary of the combined results and rescaled one-sigma error bars is presented in Table \ref{fit_table}.

Even though the quality of our spectra allows for little redundancy in the determination of the rotational constants $B$, our assignment of vibrational levels is quite solid. Observed isotope shifts are consistent with the assigned values of $\Delta v$, and the redundant pair of measurements for the $B''$ for X1.5 $v=0$ is consistent to a difference of $< 0.001$ $\wn$. The two bands with $\Omega'=3/2$ and $v=0$ permit a precise determination of $\Delta G''_{1/2} (668.641(7)$ $\wn)$ for the X1.5 state, and the two bands with $\Omega'=1/2$ and $v=1$ give a redundant, if less precise determination of the same value. The combined result is in good agreement with, but much more precise than, the previous result $\Delta G''_{1/2} = 670(13)$ $\wn$ reported by Adam et al\cite{Adam}.

We also observe one transition from the $v=2$ level of the X1.5 state. From this we are able to report a value $\Delta G''_{3/2}=663.082(9)$ $\wn$. This allows us to extract the anharmonicities $\omega_e x_e$ of the X1.5 and the $[13.8]0.5$ states, which are 2.78(1) $\wn$ and 2.6(1) $\wn$, respectively. 
Our measured value of the ground state $\omega_e x_e$ is consistent with the calculation based on the functional form of a Morse oscillator\cite{Herzberg}, $\omega_e^2/4 D_e \sim 2.0$ $\wn$, calculated from the measured bond-dissociation energy $D_e = 54000$ $\wn$\cite{dissociation}. A summary of the molecular constants for different vibrational levels of the states X1.5, $[13.8]0.5$, and $[14.2]1.5$ is in Table \ref{mol_table}.

We are not aware of published theoretical predictions for HfF, but the authors of Ref. \citenum{HfCl} provide predictions for the isoelectronic species HfCl. They predict three clusters of states: (i) three low-lying doublet states including a ground state  $^{2}\Delta_{3/2}$, (ii) a collection of quartet states around $9000$ $\wn$, and (iii) a collection of doublet states around $18000$ $\wn$. In the experimental work on HfF, the authors of Ref. \citenum{Adam}, having worked from the assumption that HfF structure will be similar to HfCl, speculated that the seven doublet levels they observed from 17000 $\wn$ to 24000 $\wn$ are part of cluster (iii). If this assumption is true, it is plausible that the $[14.2]1.5$ and $[13.8]0.5$ levels we have discussed are derived from the nominally quartet levels of group (ii).

\section{Conclusion}

We extend observations of the spectrum of HfF using LIF on a supersonic beam. After recording a survey spectrum from 13500 to 14500 $\wn$ we acquired eight vibrational bands at higher resolution. We perform rotational analysis on these bands, assigning them as  vibrational subbands of two new electronic states of HfF, $[13.8]0.5$ and $[14.2]1.5$. From the analysis we are also able to extract rotational constants for the electronic states, as well as $\Delta G''_{1/2}$ and  $\omega_e \chi_e$ for the ground state.

\section*{Acknowledgments}
We would like to thank J. Ye for the loan of a fast PMT gating circuit, and acknowledge useful conversations with P. Bernath and R. Field. This work was supported by the NSF, the  W.M. Keck Foundation, and the Marsico Endowed Chair. H. Loh acknowledges support from A*STAR (Singapore). 

\bibliography{HfF_LIF_paper}

\begin{thebibliography}{10}
\expandafter\ifx\csname url\endcsname\relax
  \def\url#1{\texttt{#1}}\fi
\expandafter\ifx\csname urlprefix\endcsname\relax\def\urlprefix{URL }\fi
\expandafter\ifx\csname href\endcsname\relax
  \def\href#1#2{#2} \def\path#1{#1}\fi

\bibitem{Leanhardt}
A.~Leanhardt, J.~Bohn, H.~Loh, P.~Maletinsky, E.~Meyer, L.~Sinclair, R.~Stutz,
  E.~Cornell, J. Mol. Spectrosc. 270~(1).

\bibitem{Meyer}
E.~R. Meyer, J.~L. Bohn, M.~P. Deskevich, Phys. Rev. A 73 (2006) 062108.

\bibitem{Barker}
B.~J. Barker, I.~O. Antonov, V.~E. Bondybey, M.~C. Heaven, J. Chem Phys. 134
  (2011) 201102.

\bibitem{Titov}
A.~N. Petrov, N.~S. Mosyagin, A.~V. Titov, Phys. Rev. A 79 (2009) 012505.

\bibitem{hinds}
J.~J. Hudson, D.~M. Kara, I.~J. Smallman, B.~E. Sauer, M.~R. Tarbutt, E.~A.
  Hinds, Nature 473 (2011) 493--496.

\bibitem{autoionization}
H.~Loh, J.~Wang, M.~Grau, T.~S. Yahn, R.~W. Field, C.~H. Greene, E.~A. Cornell,
  J. Chem. Phys. 135 (2011) 154308.

\bibitem{Adam}
A.~Adam, W.~Hopkins, D.~Tokaryk, J. Mol. Spectrosc. 225 (2004) 1--7.

\bibitem{rempi}
H.~Loh, R.~P. Stutz, T.~S. Yahn, H.~Looser, E.~A. Cornell, in preparation.

\bibitem{Cossel}
K.~C. Cossel, L.~C. Sinclair, T.~Coffey, R.~W. Field, A.~Titov, A.~Petrov,
  E.~A. Cornell, J.~Ye, in preparation.

\bibitem{Sinclair}
L.~C. Sinclair, K.~C. Cossel, T.~Coffey, J.~Ye, E.~A. Cornell, Phys. Rev. Lett.
  107 (2011) 093002.

\bibitem{BandC}
J.~M. Brown, A.~Carrington, Rotational spectroscopy of diatomic molecules,
  Cambridge University Press, 2003.

\bibitem{iupac}
K.~J.~R. Rosman, P.~D.~P. Taylor, Pure and Appl. Chem. 70~(1) (1998) 217--235.

\bibitem{bootstrap}
B.~Efron, R.~J. Tibshirani, An Introduction to the Bootstrap, Chapman and Hall,
  1994.

\bibitem{Herzberg}
G.~Herzberg, Molecular Spectra and Molecular Structure: Spectra of Diatomic
  Molecules, Krieger, 1989.

\bibitem{dissociation}
N.~V. Barkowskii, V.~I. Tsirel'nikov, A.~Emel'yanov, Y.~S. Khodeev, High
  Temperature 29~(371).

\bibitem{HfCl}
R.~Ram, A.~Adam, A.~Tsouli, J.~Li{\'e}vin, P.~Bernath, J. Mol. Spectrosc. 202
  (2000) 116--130.

\end{thebibliography}
\bibliographystyle{elsarticle-num}

\end{document}